\documentclass[aps,pre,twocolumn,showpacs,preprintnumbers,amsmath,amssymb]{revtex4}
\usepackage{epsfig}
\begin{document}
\title{Statistical analysis of 22 public transport networks in Poland}
\author{Julian Sienkiewicz and Janusz A. Ho{\l}yst}
\affiliation{Faculty of Physics and Center of Excellence for
Complex Systems Research, Warsaw University of Technology,
Koszykowa 75, PL-00-662 Warsaw, Poland}
\date{\today}

\begin{abstract}
Public transport systems in 22 Polish cities have been analyzed.
Sizes of these networks range from $N=152$ to $N=2881$. Depending
on the assumed definition of  network topology the degree
distribution can follow a power law or can be described by an
exponential function. Distributions of path lengths in all
considered networks  are given by asymmetric, unimodal functions.
Clustering, assortativity and betweenness  are studied. All
considered networks exhibit small world behavior and are
hierarchically organized. A transition between dissortative small
networks $N \lesssim 500$ and  assortative large networks $N
\gtrsim 500$ is observed.
\end{abstract} \pacs{89.75.-k, 02.50.-r, 05.50.+q} \maketitle

\section{Introduction}

Since the explosion of the complex network science that has taken
place after works of  Watts and Strogatz \cite{watts_nat} as well
as Barab\'{a}si and Albert \cite{barabasi_sci, bamf} a lot of
real-world networks  have been examined. The examples are
technological networks (Internet, phone calls network), biological
systems  (food webs, metabolic systems) or social networks
(co-authorship, citation networks) \cite{barabasi_rmp,
newman_siam, mendes_book, satorras_book}. Despite this, at the
beginning little attention has been paid to {\it transportation
networks} - mediums as much important and also sharing as much
complex structure as those previously listed. However, during the
last few years several {\it public transport systems} (PTS) have
been investigated using various concepts of statistical physics of
complex networks
\cite{amaral_pnas,strogatz_nat,albert_pre,crucitti_physa,marichiori_physa,latora_prl,latora_physa,sen_pre,seaton_physa,guimera_arxiv,
guimera_epjb,barrat_pnas,li_pre,bagler_arxiv}.

Chronogically the first works regarding transportation networks
have dealt with power grids
\cite{barabasi_sci,amaral_pnas,watts_nat,strogatz_nat}. One can
argue that transformators and transmission lines have little in
common with PTS (i.e. underground, buses and tramways), but they
definitely share at least one common feature: embedding in a
two-dimensional space. Research done on the electrical grid in
United States - for Southern California
\cite{barabasi_sci,amaral_pnas,watts_nat,strogatz_nat} and for the
whole country \cite{albert_pre} as well as on the GRTN Italian
power network \cite{crucitti_physa} revealed a single-scale degree
distributions ($p(k) \propto \exp(-\alpha k)$ with $\alpha \approx
0.5$), a small average connectivity values and relatively large
average path lengths.

All railway and underground systems appear to share well known
small-world properties \cite{watts_nat}. Moreover this kind of
networks possesses  several other characteristic features. In fact
Latora and Marichiori have studied in details  a network formed by
the Boston subway \cite{marichiori_physa, latora_prl,
latora_physa}. They have calculated  a network {\it efficiency}
defined as a mean value of inverse distances between network
nodes. Although the global efficiency is quite large
$E_{glob}=0.63$ the local efficiency calculated in the subgraphs
of neighbors is low $E_{local}=0.03$ what indicates a large
vulnerability of this network against accidental damages. However,
the last parameter increases  to $E_{local}^{'}=0.46$ if the
subway network is extended by the existing bus routes network.
Taking into account geographical distances between different metro
stations one can consider the network as a weighted graph and one
is able to introduce a measure of a network cost. The estimated
relative cost of the Boston subway is around 0.2 \% of the total
cost of fully connected network.

Sen et al. \cite{sen_pre} have introduced   a new topology
describing the system as a set of train lines, not stops, and they
have discovered a clear exponential degree distribution in Indian
railway network. This system has shown  a small negative value of
assortativity coefficient. Seaton and Hackett \cite{seaton_physa}
have compared real data from underground systems of Boston (first
presented in \cite{latora_physa}) and Vienna with the prediction
of bipartite graph theory (here: graph of lines and graph of
stops) using generation function formalism. They have found a good
correspondence regarding value of average degree, however other
properties like clustering coefficient or network size have shown
differences of 30 to 50 percent.

In  works of Amaral, Barrat, Guimer\`{a} et al. \cite{amaral_pnas,
guimera_arxiv, guimera_epjb, barrat_pnas} a survey on the
World-Wide Airport Network has been presented. The authors have
proposed truncated power-law cumulative degree distribution $P(k)
\propto k^{-\alpha}f(k/k_x)$ with the exponent $\alpha=1.0$ and a
model of preferential attachment where a new node (flight) is
introduced with a probability given by a power-law or an
exponential function of physical distance between connected nodes.
However, only an introduction of geo-political constrains
\cite{barrat_pnas} (i.e. only large cities are allowed to
establish international connections) explained the behavior of
betweenness as a function of node degree.

Other works on airport networks in India \cite{li_pre} and China
\cite{bagler_arxiv} have stressed small-world properties of those
systems, characterized by small average path lengths ($\langle l
\rangle \approx 2$) and large clustering coefficients ($c > 0.6$)
with comparison to random graph values. Degree distributions have
followed either a power-law (India) or a truncated power-law
(China). In both cases an evidence of strong disassortative
degree-degree correlation has been discovered and it also appears
that Airport Network of India has a hierarchical structure
expressed by a power-law decay of clustering coefficient with an
exponent equal to $1$.

In the present paper we have studied a part of data for PTS in
$22$ Polish cities and we have analyzed their nodes degrees, path
lengths, clustering coefficients, assortativity and betweenness.
Despite large differences in sizes of considered networks  (number
of nodes ranges from $N=152$ to $N=2881$) they share several
universal features such as degree and path length distributions,
logarithmic dependence of distances on nodes degrees or a power
law decay of clustering coefficients for large nodes degrees. As
far as we know, our results are the first comparative survey of
several public transport systems in the same country using
universal tools of complex networks.

\section{The idea of space L and P}\label{sec:deflp}

\begin{figure}[h]
 \centerline{\epsfig{file=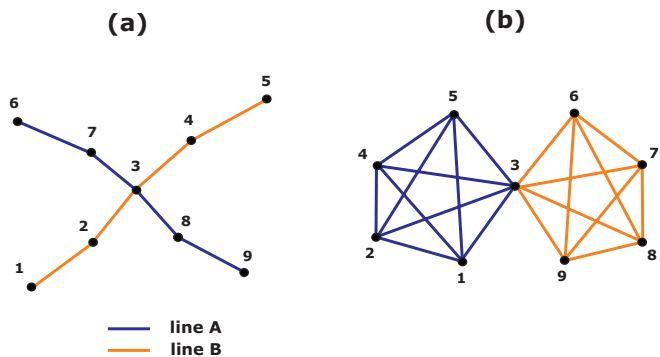,width=\columnwidth}}
    \caption{(Color online) Explanation of the space L (a) and the space P (b).}
    \label{fig:przestrzenie}
\end{figure}

To analyze various properties of PTS one should start with a
definition of a proper network topology. The idea of the space L
and P, proposed in a general form in \cite{sen_pre} and used also
in \cite{seaton_physa} is presented at Fig.
\ref{fig:przestrzenie}. The first topology (space L) consists of
nodes representing bus, tramway or underground stops and a link
between two nodes exists if they are consecutive stops on the
route. The node degree $k$ in this topology is just the number of
directions (it is usually twice the number of all PTS routes) one
can take from a given node while the distance $l$ equals to the
total number of stops on the path from one node to another.

Although nodes in the space P are the same as in the previous
topology, here an edge between two nodes means that there is a
direct bus, tramay or underground route that links them. In other
words, if a  route $A$ consists of nodes $a_i$, i.e. $A = {\{a_1,
a_2, ... , a_n\}}$, then in the space P the nearest neighbors of
the node $a_1$ are $a_2, a_3,..., a_n$. Consequently the node
degree $k$ in this topology is the total number of nodes reachable
using a single route and the distance can be interpreted as a
number of transfers (plus one) one has to take to get from one
stop to another.

\begin{figure}[ht]
 \centerline{\epsfig{file=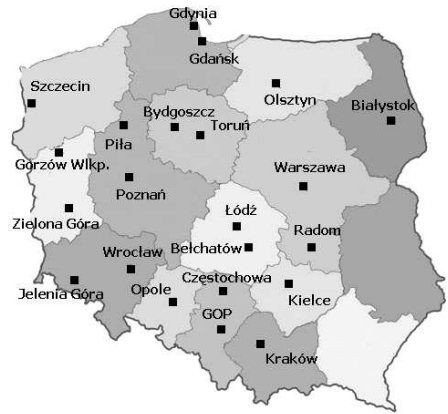,width=.7\columnwidth}}
    \caption{Map of examined cities in Poland.}
    \label{fig:mapa}
\end{figure}

Another idea of mapping a structure embedded in two-dimensional
space into another, dimensionless topology has recently  been used
by Rosvall {\it et al.} in \cite{rosvall_prl} where a plan of the
city roads has been mapped into an "information city network". In
the last topology a road represents a node and an intersection
between roads - an edge, so the network shows information handling
that has to be performed to get oriented in the city.

We need to stress that the spaces L and P do not take into account
Euclidean distance between nodes. Such an approach is similar to
the one used for description of several other types of network
systems: Internet \cite{barabasi_sci}, power grids
\cite{albert_pre,crucitti_physa}, railway \cite{sen_pre} or
airport networks \cite{li_pre,bagler_arxiv}.

\section{Explored systems}

\begin{table*}
\setlength{\tabcolsep}{6.5pt}
\begin{center}
\begin{tabular}{l|ccc|cccc|cccc}
\hline\hline
 & \multicolumn{3}{c|}{basic parameters} & \multicolumn{4}{c|}{space L} & \multicolumn{4}{c}{space P} \\
 \cline{2-12}
city & N & S & I & $ \langle l \rangle_L$ & $\langle k \rangle_L$
& $c_L$ & $r_L$ & $\langle l \rangle_P$ &
$\langle k \rangle_P$ & $c_P$ & $r_P$ \\
\hline
Pi{\l}a & 152 & 103 & 77 & 7.86 & 2.90 & 0.143 & 0.236 & 1.82 & 38.68 & 0.770 & 0.022\\
Be{\l}chat\'ow & 174 & 35 & 65 & 16.94 & 2.62 & 0.126 & 0.403 & 1.71 & 49.92 & 0.847 & -0.204\\
Jelenia G\'ora & 194 & 109 & 93 & 11.14 & 2.53 & 0.109 & 0.384 & 2.01 & 32.94 & 0.840 & 0.000\\
Opole & 205 & 96 & 129 & 10.29 & 3.03 & 0.161 & 0.320 & 1.80 & 50.19 & 0.793 & -0.108\\
Toru\'n & 243 & 116 & 206 & 10.24 & 2.72 & 0.134 & 0.068 & 2.12 & 35.84 & 0.780 & -0.055\\
Olsztyn & 268 & 88 & 173 & 12.02 & 3.08 & 0.111 & 0.356 & 1.91 & 52.91 & 0.724 & 0.020\\
Gorz\'ow Wlkp. & 269 & 77 & 162 & 16.41 & 2.48 & 0.082 & 0.401 & 2.40 & 38.51 & 0.816 & -0.033\\
Bydgoszcz & 276 & 174 & 386 & 10.48& 2.61 & 0.094 & 0.147 & 2.10 & 33.13 & 0.799 & -0.068\\
Radom & 282 & 112 & 232 & 10.97 & 2.84 & 0.089 & 0.348 & 1.98 & 48.14 & 0.786 & -0.067\\
Zielona G\'ora & 312 & 58 & 119 & 6.83 & 2.97 & 0.067 & 0.237 & 1.97 & 44.77 & 0.741 & -0.115\\
Gdynia & 406 & 136 & 255 & 11.41 & 2.78 & 0.153 & 0.307 & 2.22 & 52.68 & 0.772 & -0.018\\
Kielce & 414 & 109 & 93 & 16.98 & 2.68 & 0.122 & 0.396 & 2.05 & 48.15 & 0.771 & -0.106 \\
Cz\c{e}stochowa & 419 & 160 & 256 & 16.82 & 2.55 & 0.055 & 0.220 & 2.11 & 57.44 & 0.776 & -0.126\\
Szczecin & 467 & 301 & 417 & 12.34 & 2.54 & 0.059 & 0.042 & 2.47 & 34.55 & 0.794 & -0.004\\
Gda\'nsk & 493 & 262 & 458 & 16.14 & 2.61  & 0.132 & 0.132 & 2.30  & 40.52 & 0.804 & -0.058\\
Wroc{\l}aw & 526 & 293 & 637 & 12.52 & 2.78 & 0.147 & 0.286 & 2.24 & 50.83 & 0.738 & 0.048 \\
Pozna\'n & 532 & 261 & 577 & 14.94 & 2.72 & 0.136 & 0.194 & 2.47 & 44.87 & 0.760 & 0.160\\
Bia{\l}ystok & 559 & 90 & 285 & 11.93 & 2.76 & 0.032 & 0.004 & 2.00 & 62.55 & 0.682 & -0.076\\
Krak\'ow & 940 & 327 & 738 & 21.52 & 2.52 & 0.106 & 0.266 & 2.71 & 47.53 & 0.779 & 0.212\\
{\L}\'od\'z & 1023 & 294 & 800 & 17.10 & 2.83 & 0.065 & 0.070 & 2.45 & 59.79 & 0.721 & 0.073\\
Warszawa & 1530 & 494 & 1615 & 19.62 & 2.88 & 0.149 & 0.340 & 2.42 & 90.93 & 0.691 & 0.093\\
GOP & 2811 & 1412 & 2100 & 19.76 & 2.83 & 0.085 & 0.208 & 2.90 & 68.42 & 0.760 & -0.039\\
\hline\hline
\end{tabular}
\end{center}
\caption{Data gathered on 22 cities in Poland. $S$ stands for the
surface occupied by the city (in $km^2$) \cite{note1}, $I$ is the
city's population in thousands of inhabitants \cite{note1} and $N$
is the number of nodes (stops) in the network. $\langle l \rangle$
is the average path length, $\langle k \rangle$ - the average
degree value, $c$ is the clustering coefficient and $r$  - the
assortativity coefficient. Indexes $L$ and $P$ stand,
consequently, for the space L and the space P. Properties of
parameters defined in spaces L and P will be discussed in sec.
\ref{sec:deg}-\ref{sec:kor}} \label{tab:all}
\end{table*}

We have analyzed   PTS (bus and tramways systems) in $22$ Polish
cities, located in various state districts as it is depicted at
Fig. \ref{fig:mapa}.  Table \ref{tab:all} gathers fundamental
parameters of considered cities and data on average path lengths,
average degrees, clustering coefficients as well as  assortativity
coefficients for  corresponding networks.

\begin{figure}[ht]
 \centerline{\epsfig{file=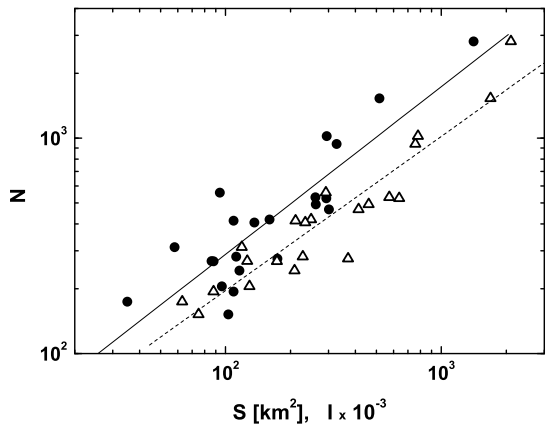,width=.9\columnwidth}}
    \caption{Log-log plot of the dependence of number of nodes $N$ on surface $S$ (circles) and population $I$ (triangles).}
    \label{fig:nlns}
\end{figure}

Numbers of nodes in different networks (i.e. in different cities)
range from $N=152$ to $N=2811$ and they are roughly proportional
to populations $I$ and surfaces $S$ of corresponding cities (see
Fig. \ref{fig:nlns}). One should notice that other surveys
exploring the properties of transportation networks have usually
dealt with smaller numbers of vertices, such as $N=76$ for U-Bahn
network in Vienna \cite{seaton_physa}, $N=79$ for Airport Network
of India (ANI) \cite{bagler_arxiv}, $N=124$ in Boston Underground
Transportation System (MBTA) \cite{latora_physa} or $N=128$ in
Airport Network of China (ANC) \cite{li_pre}. Only in the case of
the Indian Railway Network (IRN) \cite{sen_pre} where $N=579$ and
World-Wide Airport Network (WAN) \cite{barrat_pnas} with $3880$
nodes sizes of networks have been similar or larger than for PTS
in Poland. Very recently, von Ferber {\it et al.}
\cite{ferber_cmp} have presented a paper on three large PTS:
D{\"u}sseldorf with $N=1615$, Berlin with $N=2952$ and Paris where
$N=4003$.

\section{Degree distributions}\label{sec:deg}

\subsection{Degree distribution in the space L}

\begin{figure}[!ht]
 \centerline{\epsfig{file=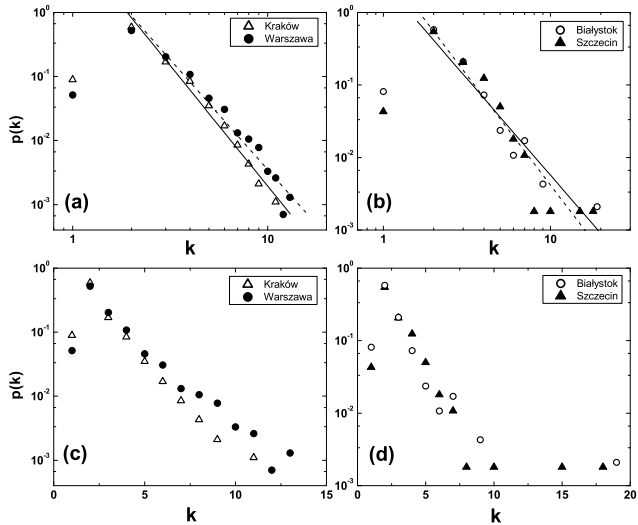,width=\columnwidth}}
    \caption{Degree distributions in the space L for four chosen cities. Plots (a) and (b) show the distributions
    in log-log scale while plots (c) and (d) - in the semi-log scale.}
    \label{fig:pl}
\end{figure}

Fig. \ref{fig:pl} shows typical plots for degree distribution in
the space L. One can see that there is a slightly better fit to
the linear behavior in the log-log description as compared to
semi-logarithmic plots. Points $k=1$ are very peculiar since they
correspond to routes' ends. Remaining parts of degree
distributions can be approximately described by a power law

\begin{equation}\label{eq:powerlaw}
p(k)\sim k^{-\gamma}
\end{equation}

although the scaling cannot be seen very clearly and it is limited
to less than one decade. Pearson correlation coefficients of the
fit to Eq. (\ref{eq:powerlaw}) range from 0.95 to 0.99. Observed
characteristic exponents $\gamma$ are between $2.4$ and $4.1$ (see
Table \ref{tab:pk}), with the majority (15 out of 22) $\gamma >
3$.  Values of exponents $\gamma$ are significantly different from
the value $\gamma=3$ which is characteristic for Barab\'asi-Albert
model of evolving networks with preferential attachment
\cite{bamf} and one can suppose that a corresponding model for
transport network evolution should include several other effects.
In fact various models taking into account effects of fitness,
atractiveness, accelerated growth and aging of vertices
\cite{mendes_adv} or deactivation of nodes
\cite{vazquez_pre,klemm_pre} lead to $\gamma$ from a wide range of
values $\gamma \in \langle 2, \infty)$. One should also notice
that networks with a characteristic exponent $\gamma> 4$ are
considered topologically close to random graphs \cite{havlin_prl}
- the degree distribution is very narrow - and a difference
between power-law and exponential behavior is very subtle (see the
Southern California power grid distribution in \cite{barabasi_sci}
presented as a power-law with $\gamma \approx 4$ and in
\cite{strogatz_nat} depicted as a single-scale cumulative
distribution).

Degree distributions obtained for airport networks are also
power-law (ANC, ANI) or power-law with an exponential cutoff (in
the case of WAN). For all those systems exponent $\gamma$ is in
the range of $2.0-2.2$, which differs significantly from
considered PTS in Poland, however one has to notice, that airport
networks are much less dependent on the two-dimensional space as
it is in the case of PTS. This effect is also seen when analyzing
average connectivity ($\langle k \rangle=5.77$ for ANI, $\langle k
\rangle=9.7$ for WAN and $\langle k \rangle = 12 - 14$ for ANC
depending on the day of the week the data have been collected).

Let us notice that the number of nodes of degree $k=1$ is smaller
as compared to the number of nodes of degree $k=2$ since $k=1$
nodes are ends of transport routes. The maximal probability
observed for nodes with degree $k=2$ means that a typical stop is
directly connected to two other stops. Still some nodes (hubs) can
have a relatively high degree value (in some cases above 10) but
the number of such vertices is very small.

\subsection{Degree distribution in the space P}

In our opinion, the key structure for the analysis of PTS are {\it
routes} and not single bus/tramway stops. Therefore we especially
take under consideration the degree distribution in the space P.

To smooth large fluctuations, we use here the cumulative
distribution $P(k)$ \cite{newman_siam} according to the formula

\begin{equation}
P(k)=\int^{k_{max}}_k p(k)dk
\end{equation}

\begin{figure}[ht]
 \centerline{\epsfig{file=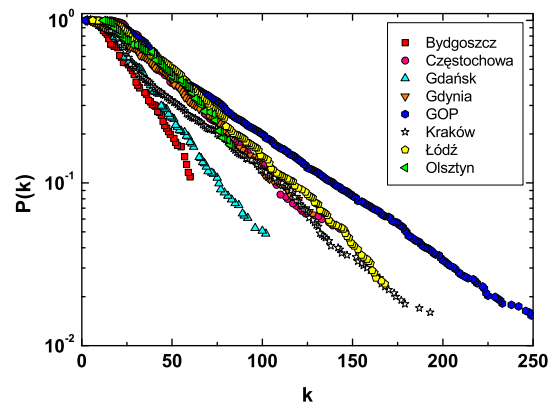,width=.9\columnwidth}}
    \caption{(Color online) $P(k)$ distribution in the space P for eight chosen cities}
    \label{fig:pk}
\end{figure}

The cumulative distributions in the space P for eight chosen
cities are shown at Fig \ref{fig:pk}. Using the semi-log scale we
observe an exponential character of such distributions:

\begin{equation}\label{eq:exponential}
P(k)=Ae^{-\alpha k}
\end{equation}

As it is well known \cite{bamf} the exponential distribution
(\ref{eq:exponential}) can occur for evolving networks when nodes
are attached completely {\it randomly}. This suggests that a
corresponding evolution of public transport in the space P
possesses an accidental character that can appear because of large
number of factors responsible for urban development. However in
the next sections we show that other network's parameters such as
clustering coefficients or degree-degree correlations calculated
for PTS are much larger as compared to corresponding values of
randomly evolving networks analyzed in \cite{bamf}.

In the case of IRN \cite{sen_pre} degree distribution in the space
P has also maintained the single-scale character $P(k) \sim e^{-\alpha
k}$ with the characteristic exponent $\alpha=0.0085$. Values of
average connectivity in the studies of MBTA ($\langle k
\rangle=27.60$) and U-Bahn in Vienna ($\langle k \rangle=20.66$)
are smaller than for considered systems in Poland, however one
should notice that sizes of networks in MBTA and Vienna are also
smaller.
\begin{table}[!ht]
\setlength{\tabcolsep}{6pt}
  \centering
  \begin{tabular}{lcccccccc}
    \hline \hline
    city & $\gamma$ & $\Delta \gamma$ & $\alpha$ & $\Delta \alpha$\\
    \hline
Pi{\l}a & 2.86 & 0.17 & 0.0310 & 0.0006\\
Be{\l}chat\'ow & 2.8 & 0.4 & 0.030 & 0.002\\
Jelenia G\'ora & 3.0 & 0.3 & 0.038 & 0.001\\
Opole & 2.29 & 0.23 & 0.0244 & 0.0004\\
Toru\'n & 3.1 & 0.4 & 0.0331 & 0.0006\\
Olsztyn & 2.95 & 0.21 & 0.0226 & 0.0004\\
Gorz\'ow Wlkp. & 3.6 & 0.3 & 0.0499 & 0.0009\\
Bydgoszcz & 2.8 & 0.3 & 0.0384 & 0.0004\\
Radom & 3.1 & 0.3 & 0.0219 & 0.0004\\
Zielona G\'ora & 2.68 & 0.20 & 0.0286 & 0.0003\\
Gdynia & 3.04 & 0.2 & 0.0207 & 0.0003\\
Kielce & 3.00 & 0.15 & 0.0263 & 0.0004\\
Cz\c{e}stochowa & 4.1 & 0.4 & 0.0264 & 0.0004\\
Szczecin & 2.7 & 0.3 & 0.0459 & 0.0006\\
Gda\'nsk & 3.0 & 0.3 & 0.0304 & 0.0006\\
Wroc{\l}aw & 3.1 & 0.4 & 0.0225 & 0.0002\\
Pozna\'n & 3.6 & 0.3 & 0.0276 & 0.0003\\
Bia{\l}ystok & 3.0 & 0.4 & 0.0211 & 0.0002\\
Krak\'ow & 3.77 & 0.18 & 0.0202 & 0.0002\\
{\L}\'od\'z & 3.9 & 0.3 & 0.0251 & 0.0001\\
Warszawa & 3.44 & 0.22 & 0.0127 & 0.0001\\
GOP & 3.46 & 0.15 & 0.0177 & 0.0002\\
\hline \hline
\end{tabular}
\caption{Coefficients  $\gamma$ and $\alpha$ with their fitting
errors $\Delta \gamma$ and $\Delta \alpha$. Fitting to the scaling
relation (\ref{eq:powerlaw}) has been performed at whole ranges of
degrees $k$ while fitting  to (\ref{eq:exponential}) has been
performed at approximately half of available ranges to exclude
large fluctuations
   occurring for higher degrees (See Fig. \ref{fig:pk}).} \label{tab:pk}
\end{table}

\subsection{Average degree and average square degree}

Taking into account the normalization condition $P(k_{min})=1$ we
get the following equations for the average degree and the average
square degree:

\begin{equation}\label{eq:k11}
    \langle k \rangle =\frac{k_{min}e^{-\alpha k_{min}}-k_{max}e^{-\alpha k_{max}}}{e^{-\alpha k_{min}}-e^{-\alpha
    k_{max}}}+\frac{1}{\alpha}
\end{equation}
\begin{eqnarray}\label{eq:k21}
\lefteqn{
    \langle k^2 \rangle=\frac{k^2_{min}e^{-\alpha k_{min}}-k^2_{max}e^{-\alpha k_{max}}}{e^{-\alpha k_{min}}-e^{-\alpha
    k_{max}}}+{}} \nonumber\\ & & {}+ \frac{2(k_{min}e^{-\alpha k_{min}}-k_{max}e^{-\alpha k_{max}})}{\alpha (e^{-\alpha k_{min}}-e^{-\alpha
    k_{max}})}+\frac{2}{\alpha^2}
\end{eqnarray}

\begin{figure}[!ht]
 \centerline{\epsfig{file=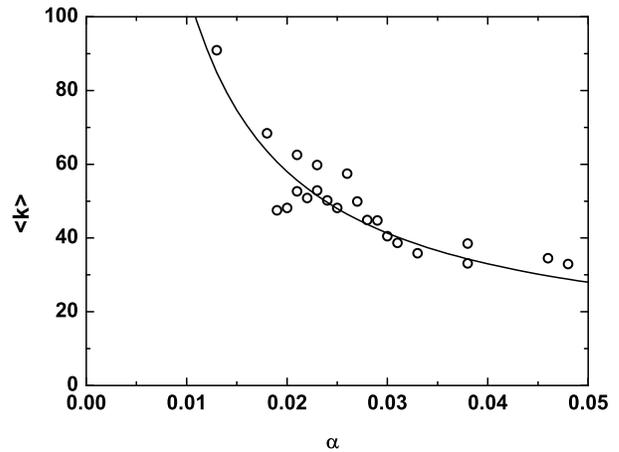,width=\columnwidth}}
    \caption{$\langle k \rangle$ as a function of $\alpha$. Circles are real data values,
    while the line corresponds to Eq. (\ref{eq:k12})}
    \label{fig:k}
\end{figure}

\begin{figure}[!ht]
 \centerline{\epsfig{file=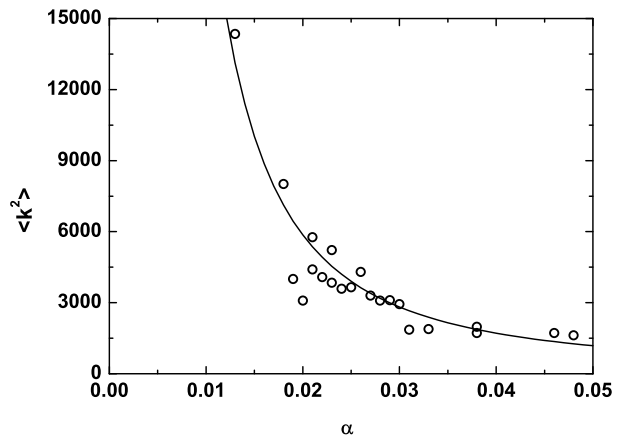,width=\columnwidth}}
    \caption{$\langle k^2 \rangle$ as a function of $\alpha$. Circles are real data values,
    while  the line corresponds to Eq. (\ref{eq:k22})}
    \label{fig:k2}
\end{figure}

Dropping all terms proportional to $e^{-\alpha k_{max}}$ we
receive simplified equations for $\langle k \rangle$ i $\langle
k^2 \rangle$:

\begin{equation}\label{eq:k12}
    \langle k \rangle \approx k_{min}+\frac{1}{\alpha}
\end{equation}
\linebreak[5]
\begin{equation}\label{eq:k22}
    \langle k^2 \rangle \approx k^2_{min}+\frac{2k_{min}}{\alpha}+\frac{2}{\alpha^2}
\end{equation}
\linebreak[5] Since values of $k_{min}$ range between $3$ and $16$
and they are independent from network sizes $N$ as well as
observed exponents $\alpha$ we have approximated $k_{min}$ in Eqs.
(\ref{eq:k12}) - (\ref{eq:k22}) by  an average value (mean
arithmetical value) for considered networks,
 $\langle k_{min}\rangle\approx 8.5$.  At Figs. \ref{fig:k} and
\ref{fig:k2} we present a comparison between the real data and
values calculated directly form Eqs. (\ref{eq:k12}) and
(\ref{eq:k22}).

\section{Path length's properties}\label{sec:path}

\subsection{Path length's distributions}\label{sec:pl}

\begin{figure}[h]
 \centerline{\epsfig{file=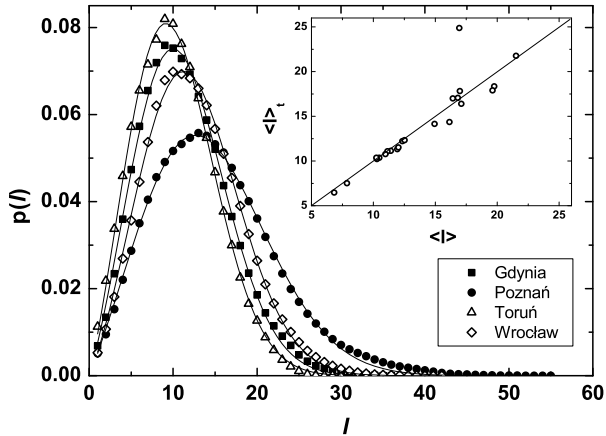,width=\columnwidth}}
     \caption{Fitted path length distribution in the space L}
    \label{fig:pll}
\end{figure}

\begin{figure}[h]
 \centerline{\epsfig{file=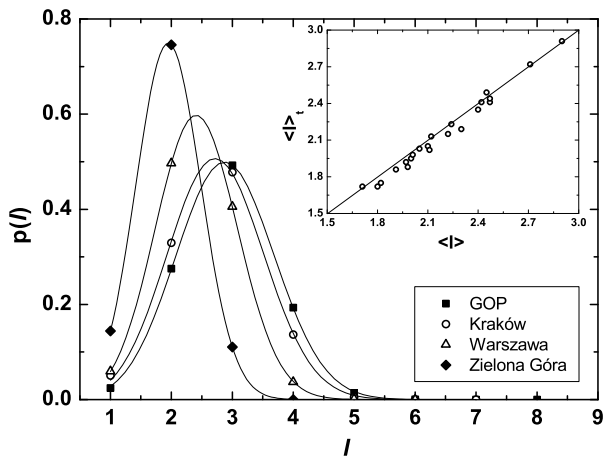,width=\columnwidth}}
     \caption{Fitted path length distribution in the space P.}
    \label{fig:plp}
\end{figure}

Plots presenting path length distributions $p(l)$ in spaces L and
P are shown at Figs. \ref{fig:pll} and \ref{fig:plp} respectively.
The data well  fit to    asymmetric, unimodal functions.  In fact
for all systems a fitting by Lavenberg - Marquardt method has been
made using the following trial function:

\begin{equation}\label{eq:pll}
    p(l) = Ale^{-Bl^2 + Cl}
\end{equation}

where $A, B$ and $C$ are fitting coefficients.

Inserts at Figs. \ref{fig:pll} and \ref{fig:plp} present a
comparison between experimental results of $\langle l \rangle$ and
corresponding mean values obtained from Eq. (\ref{eq:pll}). One
can observe a very good agreement between averages from Eq.
(\ref{eq:pll}) and experimental data.

The agreement is not surprising in the case of Fig. \ref{fig:plp}
since the number of fitted data points to curve (\ref{eq:pll}) is
quite small, but it is more prominent for Fig. \ref{fig:pll}.

Ranges of distances in the space L are much broader as compared to
corresponding ranges in the space P what is a natural effect of
topology differences. It follows that the average distance in the
space P is much smaller ($\langle l \rangle < 3$) than in the
space L. The characteristic length 3 in the space P means that in
order to travel between two different points one needs in average
no more than two transfers. Other PTS also share this property,
depending on the system size the following results have been
obtained: $\langle l \rangle=1.81$ (MBTA), $\langle l
\rangle=1.86$ (Vienna), $\langle l \rangle=2.16$ (IRN). In the
case of the space L the network MBTA with its average shortest
path length $\langle l \rangle=15.55$ is placing itself among the
values acquired for PTS in Poland. Average path length in airport
networks is  very small: $\langle l \rangle=2.07$ for ANC,
$\langle l \rangle=2.26$ for ANI and $\langle l \rangle=4.37$ for
WAN. However, because flights are usually {\it direct} (i.e. there
are no stops between two cities) one sees immediately that the
idea of the space L does not apply to airport networks - they
already have an intrinsic topology similar to the space P. Average
shortest path lengths $\langle l \rangle$ in those systems should
be relevant to values obtained for other networks after a
transformation to the space P.

The shape of path length distribution can be explained in the
following way: because transport networks tend to have an
inhomogeneous structure, it is obvious that distances between
nodes lying on the suburban routes are quite large and such a
behavior gives the effect of observed long tails in the
distribution. On the other hand shortest distances between  stops
not belonging  to suburban routes are more random  and they follow
the Gaussian distribution. A combined distribution has an
asymmetric shape with a long tail for large paths.

We need to stress that inter-node distances calculated in the
space L are much smaller as compared to the number of network
nodes (see Table I). Simultaneously clustering coefficients $c_L$
are in the range $\langle 0.03, 0.15 \rangle$. Such a behavior is
typical for small-world networks \cite{watts_nat} and  the effect
has been also observed in other transport networks
\cite{amaral_pnas,latora_physa,sen_pre,seaton_physa,li_pre,bagler_arxiv}
. The small world property is even more visible in the space P
where average distances are between $\langle 1.80, 2.90 \rangle$
and the clustering coefficient $c_P$ ranges from 0.682 to 0.847
which is similar to MBTA ($c=0.93$), Vienna ($c=0.95$) or IRN
($c=0.69$).

\subsection{Path length as function of product $k_ik_j$}\label{sec:l}

In \cite{agatac} an analytical estimation of average path length
$\langle l \rangle$ in random graphs has been found. It has been
shown that $\langle l \rangle$ can be expressed as a function of
the degree distribution. In fact the mean value for shortest path
length between $i$ and $j$ can be written as \cite{agatac}:

\begin{equation}\label{eq:lij}
    l_{ij}(k_i,k_j) = \frac{-\ln k_ik_j+\ln
    \left(\langle k^2 \rangle - \langle k \rangle \right)+\ln N - \gamma}{\ln
    \left(\langle k^2 \rangle / \langle k \rangle-1 \right)}+\frac{1}{2}
\end{equation}

where $\gamma = 0.5772$ is Euler constant.

\begin{figure}[!ht]
 \centerline{\epsfig{file=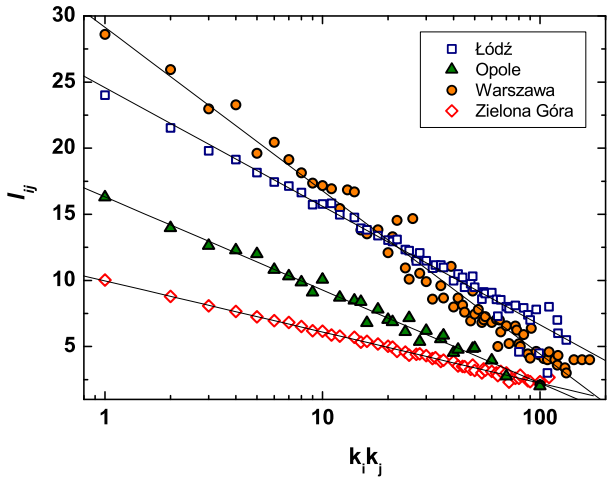,width=\columnwidth}}
     \caption{(Color online) Dependence of $l_{ij}$ on $k_ik_j$ in the space L}
    \label{fig:lij}
\end{figure}

Since PTS are not random graphs and large degree-degree
correlation in such networks exist we have assumed that Eq.
(\ref{eq:lij}) is only partially valid and we have written it a
more general form \cite{nasz_physa, nasz_prl, nasz_aip, nasz_app}:

\begin{equation}\label{eq:lijmoj}
    \langle l_{ij} \rangle = A -B \log k_ik_j.
    \end{equation}

To check the validity of Eq. (\ref{eq:lijmoj}) we have calculated
values of average path length between $l_{ij}$ as a function of
their degree product $k_ik_j$ for all systems in the space L . The
results are shown at Fig. \ref{fig:lij}, which confirms the
conjunction (\ref{eq:lijmoj}). A similar agreement has been
received for the majority of investigated PTS. Eq.
(\ref{eq:lijmoj}) can be justified using a simple model of random
graphs and a generating function formalism \cite{motter} or a
branching tree approach \cite{nasz_physa, nasz_prl, nasz_aip,
nasz_app}. In fact the scaling relation (\ref{eq:lijmoj}) can be
also observed for several other real world networks
\cite{nasz_physa, nasz_prl, nasz_aip, nasz_app}.

It is useless to examine the relation (\ref{eq:lijmoj}) in the
space P because corresponding sets $l_{ij}$ consist usually of 3
points only.

\section{Clustering coefficient}\label{sec:cluster}

We have studied clustering coefficients $c_i$ defined as a
probability that two randomly chosen neighbors of node $i$ possess
a common link.

The clustering coefficient of the whole network seems to depend
weakly on  parameters of the space L and of the space P. In the
first case its behavior with regard to network size can be treated
as fluctuations, when in the second one it is possible to observe
a small decrease of $c$ along with the networks size (see Table
\ref{tab:all}). We shall discuss only properties of the clustering
coefficients in the space P since the data in the space L are
meaningless.

\begin{figure}[!ht]
 \centerline{
 \epsfig{file=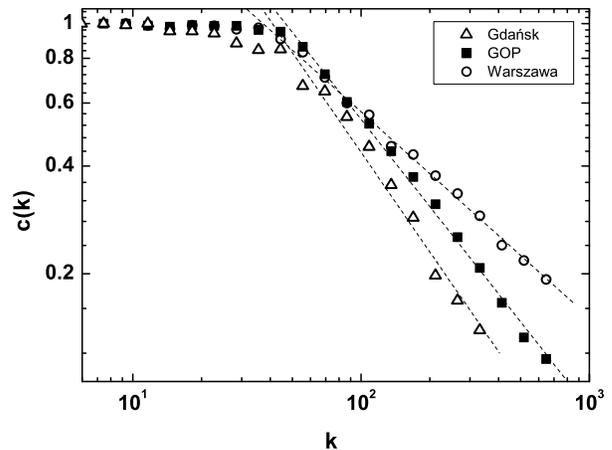,width=\columnwidth}}
 \caption{$c(k)$ for Gda\'nsk (triangles) GOP (squares) and Warszawa (circles). Dashed lines are fits to
 Eq. (\ref{eq:ck}) with following exponents: Gda\'nsk - $\beta = 0.93 \pm 0.05$, GOP - $\beta = 0.81 \pm 0.02$
 and Warszawa - $\beta = 0.57 \pm 0.01$. All data are logarithmically binned with the power of 1.25.}
 \label{fig:ckmiasta}
\end{figure}

It has been shown in \cite{sen_pre} that clustering coefficient in
IRN in the space P decays linearly with the logarithm of degree
for large $k$ and is almost constant (and close to unity) for
small $k$. In the considered PTS we have found that this
dependency can be described by a power law (see Fig.
\ref{fig:ckmiasta}):

\begin{equation}\label{eq:ck}
    c(k) \sim k^{-\beta}
\end{equation}

Such a behavior has been observed in many real systems with
hierarchical structures \cite{ravasz_pre,ravasz_sci}. In fact, one
can expect that PTS should consist of densely connected modules
linked by longer paths.

Observed values of exponents $\beta$ are in the range $\beta \in
\langle 0.54, 0.93 \rangle$. This can be explained using a simple
example of a {\it star} network: suppose that the city transport
network is a star consisting of $n$ routes with $L$ stops each.
Node $i$, at which all $n$ routes cross is a vertex that has the
highest degree in the network. We do not allow any other crossings
among those $n$ routes in the whole system. It follows that the
degree of node $i$ is $k_i=n(L-1)$ and the total number of links
among the nearest neighbors of $i$ is $E_i=n(L-1)(L-2)/2$. In
other words the value of the clustering coefficient for the node
with the maximum degree is:

\begin{equation}\label{eq:cmax}
    c(k_{max}) = \frac{2E_i}{k_i(k_i-1)}=\frac{L-2}{n(L-1)-1}
\end{equation}

where $k_{max}=n(L-1)$. It is obvious that the minimal degree in
the network is $k_{min}=L-1$ and this correspondences to the value
$c(k_{min})=1$. Using these two points and assuming that we have a
power-law behavior we can express $\beta$ as:

\begin{equation}\label{eq:beta1}
    \beta = -\frac{\ln c(k_{max}) - \ln
    c(k_{min})}{\ln k_{max} - \ln
    k_{min}}=-\frac{\ln\frac{L-2}{n(L-1)-1}}{\ln n}
\end{equation}

Because $n(L-1) \gg 1$ and $L-1 \approx L-2$ we have
    $\beta \approx 1$.

In real systems the value of clustering coefficient of the highest
degree node is larger than in Eq. (\ref{eq:cmax}) due to multiple
crossings of routes in the whole network what leads to a decrease
of the exponent $\beta$ (see Fig. \ref{fig:ckmiasta}). This
decrease is also connected to the presence of degree-degree
correlations (see the next Section).

\section{Degree-degree correlations}\label{sec:kor}

To analyze  degree-degree correlations in PTS we have used the
assortativity coefficient $r$, proposed by Newman \cite{new2} that
corresponds to the Pearson correlation coefficient \cite{new4} of
the nodes degrees at the end-points of link:

\begin{equation}\label{eq:rsimp}
    r = \frac{\sum_i j_ik_i - \frac{1}{M}\sum_i j_i \sum_i k_i}{\sqrt{\sum_i j_i^2 - \frac{1}{M}(\sum_i j_i)^2}\sqrt{\sum_i k_i^2 - \frac{1}{M}(\sum_i k_i)^2}}
\end{equation}

where $M$ - number of pairs of nodes (twice the number of edges),
$j_i, k_i$  - degrees of vertices at both ends of $i$-th pair and
index $i$ goes over all pairs of nodes in the network.

\begin{figure}[!ht]
 \centerline{\epsfig{file=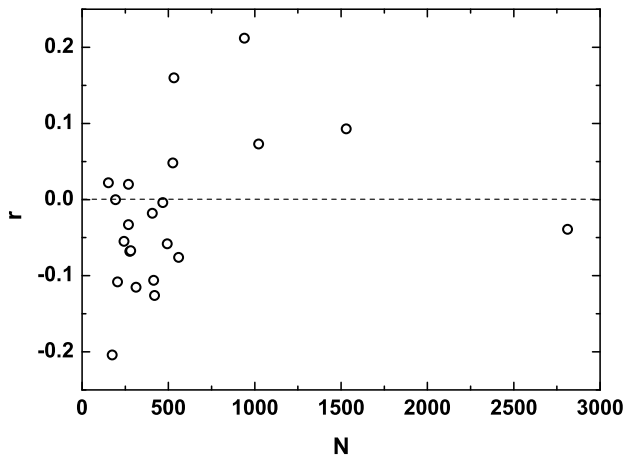,width=\columnwidth}}
     \caption{The assortativity coefficient $r$ in the space P as a function of $N$}
    \label{fig:rp}
\end{figure}

Values of the assortativity coefficient $r$ in the space L are
independent of the network size and are always positive (see Table
\ref{tab:all}), what can be explained in the following way: there
is a little number of nodes characterized by high values of
degrees $k$ and they are usually linked among themselves. The
majority of remaining links connect nodes of degree $k=2$ or
$k=1$, because $k=2$ is an overwhelming degree in networks.

\begin{figure}[!ht]
 \centerline{
 \epsfig{file=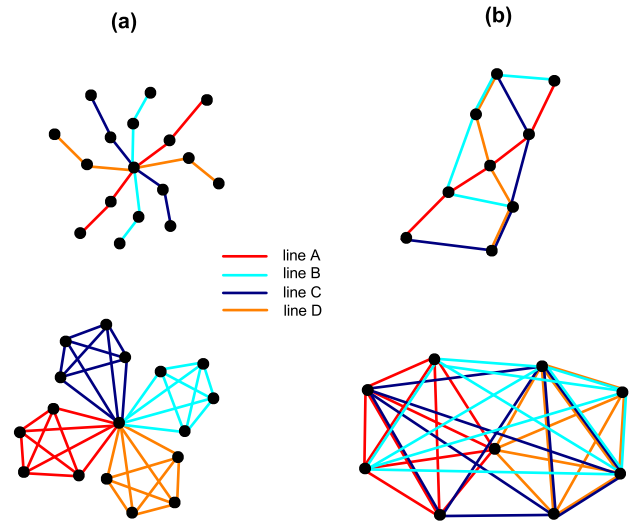,width=\columnwidth}}
 \caption{(Color online) Crossing of 4 routes of 5 stops each. (a) In the {\it star} example there is only one {\it hub} and assortativity coefficient is equal to
 $r=-0.25$ according to Eq. (\ref{eq:rstar}). In case (b) a few hubs exist due to a multiple crossing of routes and $r=-0.19$. Upper
 diagrams - the space L, lower diagrams - the space P.}
 \label{fig:struct}
\end{figure}

Similar calculations performed for the space P lead to completely
different results (Fig. \ref{fig:rp}). For small networks the
correlation parameter $r$ is negative and it grows with $N$,
becoming positive for $N \gtrsim 500$. The dependence can be
explained as follows: small towns are described by star structures
and there are only a few {\it doubled routes}, so in this space a
lot of links between vertices of small and large $k$ exist. Using
the previous example of a star network and taking into account
that the degree of the central node is equal to $k_c=n(L-1)$, the
degree of any other node is $k_o=L-1$, after some algebra we
receive the following expression for the assortativity coefficient
of such a star network:

\begin{equation}\label{eq:rstar}
    r = -\frac{1}{L-1}
\end{equation}

Let us notice that the coefficient $r$ is independent from the
number of crossing routes and is always a negative number.

On the contrary, in the large cities there are lots of connections
between nodes characterized by large $k$ (transport hubs) as well
as there is a large number of routes crossing in more than one
point (see Fig. \ref{fig:struct}). It follows that the coefficient
$r$ can be positive for such networks. A strange behavior for the
largest network (GOP) can be explained as an effect of its
peculiar structure: the system is rather a conglomerate of many
towns than a single city. Thus, the value of $r$ is lowered by
single links between the subsets of this network.

\begin{figure}[!ht]
 \centerline{
 \epsfig{file=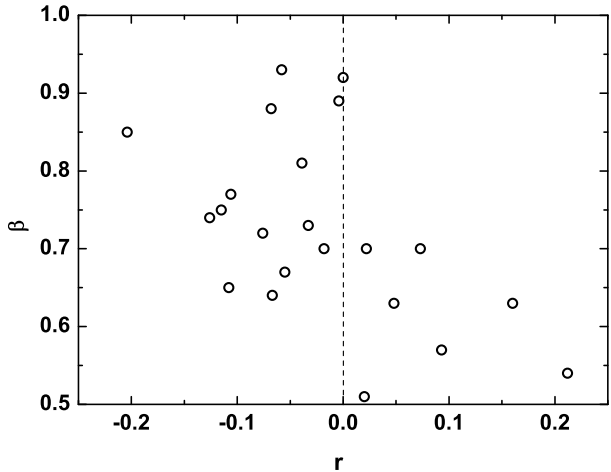,width=\columnwidth}}
 \caption{$\beta$ coefficient (see Eq. (\ref{eq:ck})) as a function of the assortativity coefficient $r$ in the space P.}
 \label{fig:betar}
\end{figure}

At Fig. \ref{fig:betar} we show coefficients $\beta$ as a function
of $r$ in the space P. One can see that in general positive values
of the assortativity coefficient correspond to lower values of
$\beta$, being an effect of existence of several links between
hubs in the networks.

Reported values of assortativity coefficients in other transport
networks  have been negative ($r = -0.402$ for ANI \cite{li_pre}
and $r = -0.033$ for IRN \cite{sen_pre}) and since  these systems
are of the size $N < 600$ thus it is in agreement with our
results.

\section{Betweenness}\label{sec:bet}

The last property of PTS examined in this work is betweenness
\cite{soc} which is the quantity  describing the "importance" of a
specific node according to equation \cite{bar1}:

\begin{equation}\label{eq:pos1}
    g(i) = \sum_{j \neq k}\frac{\sigma_{jk}(i)}{\sigma_{jk}}
\end{equation}

where, $\sigma_{jk}$ is a number of the shortest paths  between
nodes $j$ and $k$, while $\sigma_{jk}(i)$ is a  number of these
paths that go through the node $i$.

\begin{figure}[!ht]
 \centerline{\epsfig{file=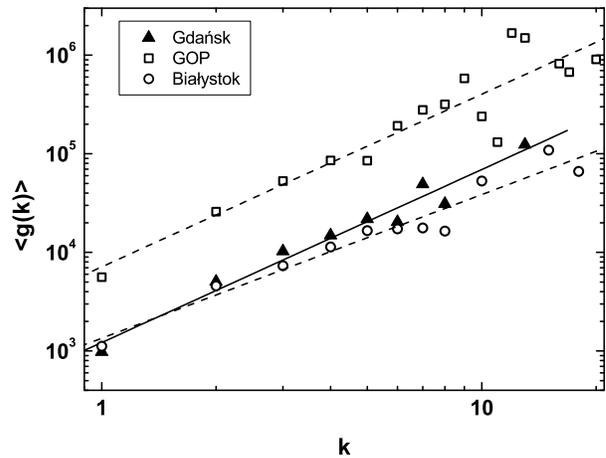,width=\columnwidth}}
     \caption{The average betweenness $\langle g\rangle$ as a function of $k$ in the space L
    for three chosen cities.}
    \label{fig:bkl}
\end{figure}

\begin{figure}
 \centerline{\epsfig{file=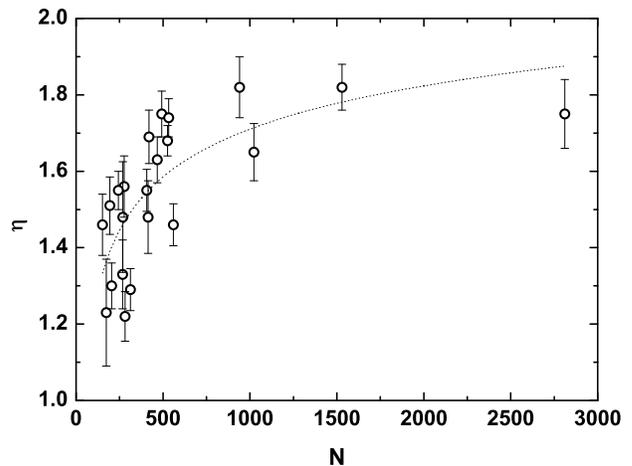,width=\columnwidth}}
     \caption{$\eta$ coefficient as a function of network size $N$}
    \label{fig:eta}
\end{figure}

\begin{figure}[!ht]
 \centerline{
 \epsfig{file=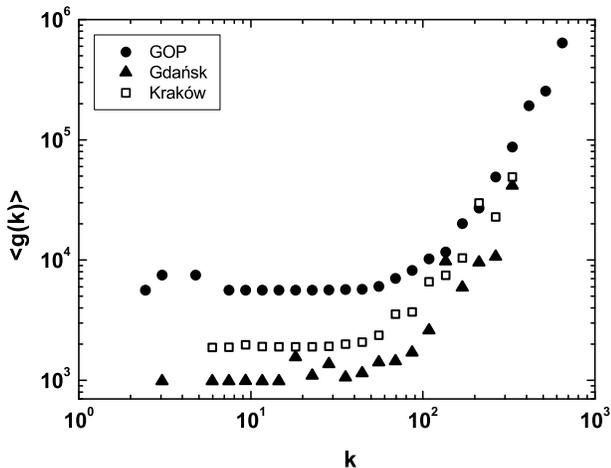,width=\columnwidth}}
     \caption{The average betweenness $\langle g\rangle$ as a function of $k$ in the space P
    for three chosen cities.}
    \label{fig:bkp}
\end{figure}

    \subsection{Betweenness in the space L}

Fig. \ref{fig:bkl} shows dependence of the average betweenness
$\langle g \rangle$ on node degree calculated using the algorithm
proposed in \cite{new3} (see also \cite{brandes}). Data at Fig.
\ref{fig:bkl} fit well to the scaling relation:

\begin{equation}\label{eq:gk}
    g \sim k^{\eta}
\end{equation}

observed in Internet Autonomous Systems \cite{vaz}, co-authorship
networks \cite{goh} and BA model or Erd\H{o}s-R\'{e}nyi random
graphs \cite{bar1}.

The coefficient $\eta$ is plotted at Fig. \ref{fig:eta} as a
function of network size. One can see, that $\eta$ is getting
closer to 2 for large networks. Since it has been shown that there
is $\eta=2$ for random graphs \cite{bar1} with Poisson degree
distribution thus it can suggest that large PTS are more random
than small ones. Such an interpretation can be also received from
the Table \ref{tab:pk} where larger values of the exponent
$\gamma$ are observed for large cities.

   \subsection{Betweenness in the space P}

The betweenness as a function of node degree $k$ in the space P is
shown at Fig. \ref{fig:bkp}. One can see large differences between
Fig. \ref{fig:bkl} and \ref{fig:bkp}. In the space P there is a
saturation of $\langle g \rangle$ for small $k$ what is a result
of existence of the suburban routes while the scale-free behavior
occurs only for larger $k$. The saturation value observed in the
limit of small $k$ is given by $\langle g(k_{min}) \rangle =
2(N-1)$ and the length of the saturation line  increases with the
mean value of a single route's length observed in a city.

\section{Conclusions}\label{sec:con}

In this study we have collected and analyzed data for public
transport networks in 22 cities that make over 25 \% of population
in Poland. Sizes of these networks range from $N=152$ to $N=2881$.
Using the concept of different network topologies we show that in
the space L, where distances are measured in numbers of passed
bus/tramway stops, the degree distributions are approximately
given by a power laws with $\gamma = 2.4 - 4.1$ while in the space
P, where distances are measured in  numbers of transfers, the
degree distribution is exponential with characteristic exponents
$\alpha=0.013 - 0.050$. Distributions of paths in both topologies
are approximately given by a function $p(l)=Ale^{-Bl^2+Cl}$. Small
world behavior is observed in both topologies but it is much more
pronounced  in space P where the hierarchical structure of network
is also deduced from the behavior of $c(k)$. The assortativity
coefficient measured in the space L remains positive for the whole
range of $N$ while in the space P it changes from negative values
for small networks to positive values for large systems. In the
space L distances between two stops are linear functions of the
logarithm of their degree products.

Many of our results are similar to features observed  in other
works regarding transportation networks: underground, railway or
airline systems
\cite{amaral_pnas,marichiori_physa,latora_prl,latora_physa,sen_pre,seaton_physa,guimera_arxiv,
guimera_epjb,barrat_pnas,li_pre,bagler_arxiv,ferber_cmp}. All such
networks tend to share small-world properties and show strong
degree-degree correlations that reveal complex nature of those
structures.

\begin{acknowledgments}
The work was supported by the EU Grant {\it Measuring and
Modelling Complex Networks Across Domains - MMCOMNET} (Grant No.
FP6-2003-NEST-Path-012999), by the State Committee for Scientific
Research in Poland (Grant No. 1P03B04727) and by a special Grant
of Warsaw University of Technology.
\end{acknowledgments}

\end{document}